\documentclass[conference]{IEEEtran}

\usepackage{graphicx}
\usepackage{epstopdf}
\usepackage{standalone}
\usepackage[nolist ]{acronym}
\usepackage{tcolorbox}

\usepackage{pdfpages}
\usepackage{tikz}
\usetikzlibrary{positioning, arrows, arrows.meta}
\usepackage[draft]{hyperref}
\usepackage{pdfcomment}
\usepackage{hologo}

\usepackage{xstring}

\usepackage[utf8]{inputenc}
\usepackage{marvosym}

\usepackage{algorithm}
\usepackage{algpseudocode}
\usepackage{tikz}

\usepackage{listings}
\definecolor{ListingBackground}{rgb}{0.97,0.97,0.97}
\usepackage{pgfplots}
\pgfplotsset{compat=newest}
\pgfplotsset{
    box plot/.style={
        /pgfplots/.cd,
        fill=blue!30,
        only marks,
        mark=-,
        mark size=0.2em,
        /pgfplots/error bars/.cd,
        y dir=plus,
        y explicit,
    },
    box plot box/.style={
        /pgfplots/error bars/draw error bar/.code 2 args={%
            \draw  ##1 -- ++(.2em,0pt) |- ##2 -- ++(-.2em,0pt) |- ##1 -- cycle;
        },
        /pgfplots/table/.cd,
        y index=2,
        y error expr={\thisrowno{3}-\thisrowno{2}},
        /pgfplots/box plot
    },
    box plot top whisker/.style={
        /pgfplots/error bars/draw error bar/.code 2 args={%
            \pgfkeysgetvalue{/pgfplots/error bars/error mark}%
            {\pgfplotserrorbarsmark}%
            \pgfkeysgetvalue{/pgfplots/error bars/error mark options}%
            {\pgfplotserrorbarsmarkopts}%
            \path ##1 -- ##2;
        },
        /pgfplots/table/.cd,
        y index=4,
        y error expr={\thisrowno{2}-\thisrowno{4}},
        /pgfplots/box plot
    },
    box plot bottom whisker/.style={
        /pgfplots/error bars/draw error bar/.code 2 args={%
            \pgfkeysgetvalue{/pgfplots/error bars/error mark}%
            {\pgfplotserrorbarsmark}%
            \pgfkeysgetvalue{/pgfplots/error bars/error mark options}%
            {\pgfplotserrorbarsmarkopts}%
            \path ##1 -- ##2;
        },
        /pgfplots/table/.cd,
        y index=5,
        y error expr={\thisrowno{3}-\thisrowno{5}},
        /pgfplots/box plot
    },
    box plot median/.style={
        /pgfplots/box plot
    },
    boxplot/every median/.style={
    	ultra thick,dashed,cyan
    }
}

\definecolor{flexicolor}{RGB}{46,49,146}
\definecolor{amaricolor}{RGB}{237,28,36}

\usepackage{xspace}

\usepackage[binary-units=true]{siunitx}
\usepackage{psfrag}
\usepackage{graphicx}
\usepackage{tabularx,booktabs}
\usepackage{multirow}
\usepackage{rotating}

\renewcommand{\baselinestretch}{1}

\usepackage{multirow}

\usepackage[cmex10]{amsmath}
\DeclareMathOperator*{\argmax}{arg\,max}

\usepackage[caption=false,font=footnotesize]{subfig}
\usepackage{placeins}

\usepackage{stfloats}
\usepackage{wasysym}
\usepackage{fontawesome}
\newcommand\full{$\CIRCLE$}
\newcommand\partly{$\LEFTcircle$}
\newcommand\missing{$\Circle$}

\newif\ifdoubleblind % https://tex.stackexchange.com/questions/33576/conditional-typesetting-build
\begin{document}

\newcommand{\paperTitle}{Towards Cooperative Data Rate Prediction for Future Mobile and Vehicular 6G Networks}
\newcommand{\paperAuthors}{Benjamin Sliwa, Robert Falkenberg, and Christian Wietfeld}
\newcommand{\paperEmails}{$\{$Benjamin.Sliwa, Robert.Falkenberg, Christian.Wietfeld$\}$@tu-dortmund.de}

\newcommand{\figurePadding}{0pt}
\newcommand{\figureTopPadding}{\figurePadding}
\newcommand{\figureBottomPadding}{\figurePadding}
\newcommand{\client}{\ac{UE}-based\xspace}
\newcommand{\net}{Network-based\xspace}
\newcommand{\coop}{Cooperative\xspace}

\newcommand{\dummy}[3]
{
	\begin{figure}[b!]  
		\begin{tikzpicture}
		\node[draw,minimum height=6cm,minimum width=\columnwidth]{\LARGE #1};
		\end{tikzpicture}
		\caption{#2}
		\label{#3}
	\end{figure}
}

\newcommand{\wDummy}[3]
{
	\begin{figure*}[b!]  
		\begin{tikzpicture}
		\node[draw,minimum height=6cm,minimum width=\textwidth]{\LARGE #1};
		\end{tikzpicture}
		\caption{#2}
		\label{#3}
	\end{figure*}
}

\newcommand{\basicFig}[7]
{
	\begin{figure}[#1]  	
		\vspace{#6}
		\centering		  
		\includegraphics[width=#7\columnwidth]{#2}
		\caption{#3}
		\label{#4}
		\vspace{#5}	
	\end{figure}
}
\newcommand{\fig}[4]{\basicFig{#1}{#2}{#3}{#4}{0cm}{0cm}{1}}

\newcommand{\subfig}[3]%
{%
	\subfloat[#3]%
	{%
		\includegraphics[width=#2\textwidth]{#1}%
	}%
	\hfill%
}%

\newcommand\circled[1] % caution with using in captions: \protect \circled
{
	\tikz[baseline=(char.base)]
	{
		\node[shape=circle,draw,inner sep=1pt] (char) {#1};
	}\xspace
}

\newcommand{\sideHeader}[3]
{
	\multirow{#1}{*}{
		\rotatebox[origin=c]{90}{
			\parbox{#2}{\centering \textbf{#3}}
		}
	}
}

\DeclareRobustCommand\boxedArrowUp{%
	\begin{tikzpicture}
%		\newlength{\mylen}
%		\setlength{\mylen}{\fontcharht\font`X}   %% height of X
		\node[draw, inner sep=0pt, outer sep=0pt] (box) {\phantom{X}};
		\draw[shorten >=0.5pt, shorten <=0.5pt, -{Triangle[angle=60:0.8ex]}] (box.south) -- (box.north);
	\end{tikzpicture}
}

\DeclareRobustCommand\boxedArrowDown{%
	\begin{tikzpicture}
%	\newlength{\mylen}
%	\setlength{\mylen}{\fontcharht\font`X}   %% height of X
	\node[draw, inner sep=0pt, outer sep=0pt] (box) {\phantom{X}};
	\draw[shorten >=0.5pt, shorten <=0.5pt, -{Triangle[angle=60:0.8ex]}] (box.north) -- (box.south);
	\end{tikzpicture}
}
\begin{acronym}
	\acro{HD}{High Definition}
	\acro{UAV}{Unmanned Aerial Vehicle}
	\acro{SUS}{System Under Study}
	\acro{AI}{Artificial Intelligence}
	\acro{CR}{Cognitive Radio}
	\acro{DDNS}{Data-driven Network Simulation}
	\acro{DES}{Discrete Event Simulation}
	\acro{MUS}{Method Under Study}
	\acro{TRUST}{Throughput prediction based on LSTM}
	\acro{LSTM}{Long Short-term Memory}
	\acro{CA}{Carrier Aggregation}
	\acro{ECDF}{Empirical Cumulative Distribution Function}
	\acro{QoS}{Quality of Service}
	\acro{URLLC}{Ultra Reliable Low Latency Communications}
	\acro{mMTC}{Massive Machine-type Communications}
	\acro{eMBB}{Enhanced Mobile Broadband}
	\acro{LIDAR}{Light Detection and Ranging}
	\acro{CAV}{Connected and Automated Vehicle}
	\acro{LR}{Linear Regression}
	\acro{TPC}{Transmission Power Control}
	\acro{AI}{Artificial Intelligence}
	
	\acro{TCP}{Transmission Control Protocol}
	\acro{KPI}{Key Performance Indicator}
	\acro{RTT}{Round Trip Time}		
	\acro{MTC}{Machine-type Communication}

	\acro{NIC}{Network Interface Card}
	\acro{PDCP}{Packet Data Convergence Protocol}
	\acro{RLC}{Radio Link Control}
	\acro{MAC}{Medium Access Control}
	\acro{HARQ}{Hybrid Automatic Repeat Request}
	\acro{IP}{Internet Protocol}

	\acro{ANN}{Artificial Neural Network}
	\acro{CART}{Classification And Regression Tree}
	\acro{GPR}{Gaussian Process Regression}
	\acro{M5}{M5 Regression Tree}
	\acro{RF}{Random Forest}
	\acro{SVM}{Support Vector Machine}
	\acro{SMO}{Sequential Minimal Optimization}
	\acro{RBF}{Radial Basis Function}
	\acro{WEKA}{Waikato Environment for Knowledge Analysis}
	\acro{MDI}{Mean Decrease Impurity}
	\acro{SGD}{Stochastic Gradient Descent}
	
	\acro{CM}{Connectivity Map}
	\acro{CAT}{Channel-aware Transmission}
	\acro{ML-CAT}{Machine Learning CAT}
	\acro{pCAT}{predictive CAT}
	\acro{ML-pCAT}{Machine Learning pCAT}
	
	\acro{ITS}{Intelligent Transportation System}
	\acro{LIMoSim}{Lightweight ICT-centric Mobility Simulation}
	\acro{OMNeT++}{Objective Modular Network Testbed in C++}
	\acro{ns-3}{Network Simulator 3}
	\acro{MAE}{Mean Absolute Error}
	\acro{RMSE}{Root Mean Square Error}
	\acro{RAIK}{Regional Analysis to Infer KPIs}
	\acro{RAT}{Radio Access Technology}
	\acro{MNO}{Mobile Network Operator}
	\acro{LTE}{Long Term Evolution}
	\acro{UE}{User Equipment}
	\acro{eNB}{evolved Node~B}
	\acro{RSRP}{Reference Signal Received Power}
	\acro{RSRQ}{Reference Signal Received Quality}
	\acro{SINR}{Signal-to-interference-plus-noise Ratio}
	\acro{CQI}{Channel Quality Indicator}
	\acro{ASU}{Arbitrary Strength Unit} 
	\acro{TA}{Timing Advance}
	\acro{NWDAF}{Network Data Analytics Function}
	\acro{FALCON}{Fast Analysis of LTE Control channels}
	\acro{C3ACE}{Client-Based Control Channel Analysis for Connectivity Estimation}
	\acro{E-C3ACE}{Enhanced C3ACE}
	\acro{OWL}{Online Watcher for LTE}
	\acro{IoT}{Internet of Things}
	\acro{SDR}{Software Defined Radio}
	\acro{3GPP}{3rd Generation Partnership Project}
	
	\acro{DCI}{Downlink Control Information}
	\acro{RNTI}{Radio Network Temporary Identifier}
	\acro{PDCCH}{Physical Downlink Control Channel}
	\acro{PRB}{Physical Resource Block}
	\acro{TBS}{Transport Block Size}
	\acro{MCS}{Modulation and Coding Scheme}
	\acro{GPR}{Gaussian Process Regression}
\end{acronym}

\acresetall
\title{\paperTitle}

\ifdoubleblind
\author{\IEEEauthorblockN{\textbf{Anonymous Authors}}
	\IEEEauthorblockA{Anonymous Institutions\\
		e-mail: Anonymous Emails}}
\else
\author{\IEEEauthorblockN{\textbf{\paperAuthors}}
	\IEEEauthorblockA{Communication Networks Institute,	TU Dortmund University, 44227 Dortmund, Germany\\
		e-mail: \paperEmails}}
\fi

\maketitle

\begin{tikzpicture}[remember picture, overlay]
\node[below=5mm of current page.north, text width=20cm,font=\sffamily\footnotesize,align=center] {Accepted for presentation in: 2nd 6G Wireless Summit (6G SUMMIT)\vspace{0.3cm}\\\pdfcomment[color=yellow,icon=Note]{
@InProceedings\{Sliwa2020towards,\\
	Author = \{Benjamin Sliwa and Robert Falkenberg and Christian Wietfeld\},\\
	Title = \{Towards cooperative data rate prediction for future mobile and vehicular \{6G\} networks\},\\
	Booktitle = \{2nd 6G Wireless Summit (6G SUMMIT)\},\\
	Year = \{2020\},\\
	Address = \{Levi, Finland\},\\
	Month = \{Mar\},\\
	Publisher = \{IEEE\},\\
\}
}};
\node[above=5mm of current page.south, text width=15cm,font=\sffamily\footnotesize] {2020~IEEE. Personal use of this material is permitted. Permission from IEEE must be obtained for all other uses, including reprinting/republishing this material for advertising or promotional purposes, collecting new collected works for resale or redistribution to servers or lists, or reuse of any copyrighted component of this work in other works.};
\end{tikzpicture}

% !TeX spellcheck = en_US
\begin{abstract}
	
%
% Introduction
%
Machine learning-based data rate prediction is one of the key drivers for anticipatory mobile networking with applications such as dynamic \ac{RAT} selection, opportunistic data transfer, and predictive caching.	
%
% Limitations of current approaches
%
\client prediction approaches that rely on passive measurements of network quality indicators have successfully been applied to forecast the throughput of vehicular data transmissions. However, the achievable prediction accuracy is limited as the \ac{UE} is unaware of the current network load.
%
% Own approach
%
To overcome this issue, we propose a cooperative data rate prediction approach which brings together knowledge from the client and network domains.
%
% Methods
%
In a real world proof-of-concept evaluation, we utilize the \ac{SDR}-based control channel sniffer FALCON in order to mimic the behavior of a possible network-assisted information provisioning within future 6G networks.  
%
% Results
%
The results show that the proposed cooperative prediction approach is able to reduce the average prediction error by up to 30\%. 
%
% 6G Recommendations
%
With respect to the ongoing standardization efforts regarding the implementation of intelligence for network management, we argue that future 6G networks should go beyond network-focused approaches and actively provide load information to the \acp{UE} in order to fuel pervasive machine learning and catalyze \client network optimization techniques.

\end{abstract}

\IEEEpeerreviewmaketitle

% !TeX spellcheck = en_US
\section{Introduction}

%
% Introduction
%
%
Although the concrete technological improvements of future 6G networks are still unclear, researches agree that data-driven intelligence will be a key driver for those novel networks which are expected to be deployed around 2030 \cite{Aazhang/etal/2019a}.
As a consequence, the \ac{3GPP} is currently investigating data analytics-based networking as an enabler for network automation \cite{3GPP/2019b}. An example is the \ac{NWDAF}, which has been specified in \cite{3GPP/2019a} as a novel 5G core network function allowing \acp{MNO} to monitor the load of network slices based on machine learning methods.

%
% Anticipatory Communication
%
While the ongoing standardization efforts for 5G mainly target \emph{network-side} intelligence, different studies have demonstrated that edge-based \cite{Park/etal/2019a} and \emph{\client} optimization is not only able to improve the end user experience, but also contributes to improving the intra-cell coexistence of different devices. \emph{Anticipatory} communication \cite{Bui/etal/2017a} has emerged as a novel mobile networking paradigm that aims to optimize decision processes by taking context information into account.
%
% Data Rate Prediction
%
In this field, machine learning-based data rate prediction is a key enabler for different applications. It allows to chose the best network interface in multi-\ac{RAT} systems \cite{Bouali/etal/2016a}, predictively cache streaming data  \cite{Mangla/etal/2016a}, and increase the resource efficiency of \ac{mMTC} through opportunistic data transfer \cite{Sliwa/etal/2019d}. Moreover, end-to-end prediction models themselves can serve as highly accurate performance analysis tools based on \ac{DDNS} \cite{Sliwa/Wietfeld/2019a} techniques. Therefore, the optimization of the achievable prediction accuracy is a crucial research task which directly affects the performance of these applications.
%
%
%

\begin{figure}[!b]  	
	\vspace{-0.7cm}
	\centering		  
	\includegraphics[width=1\columnwidth]{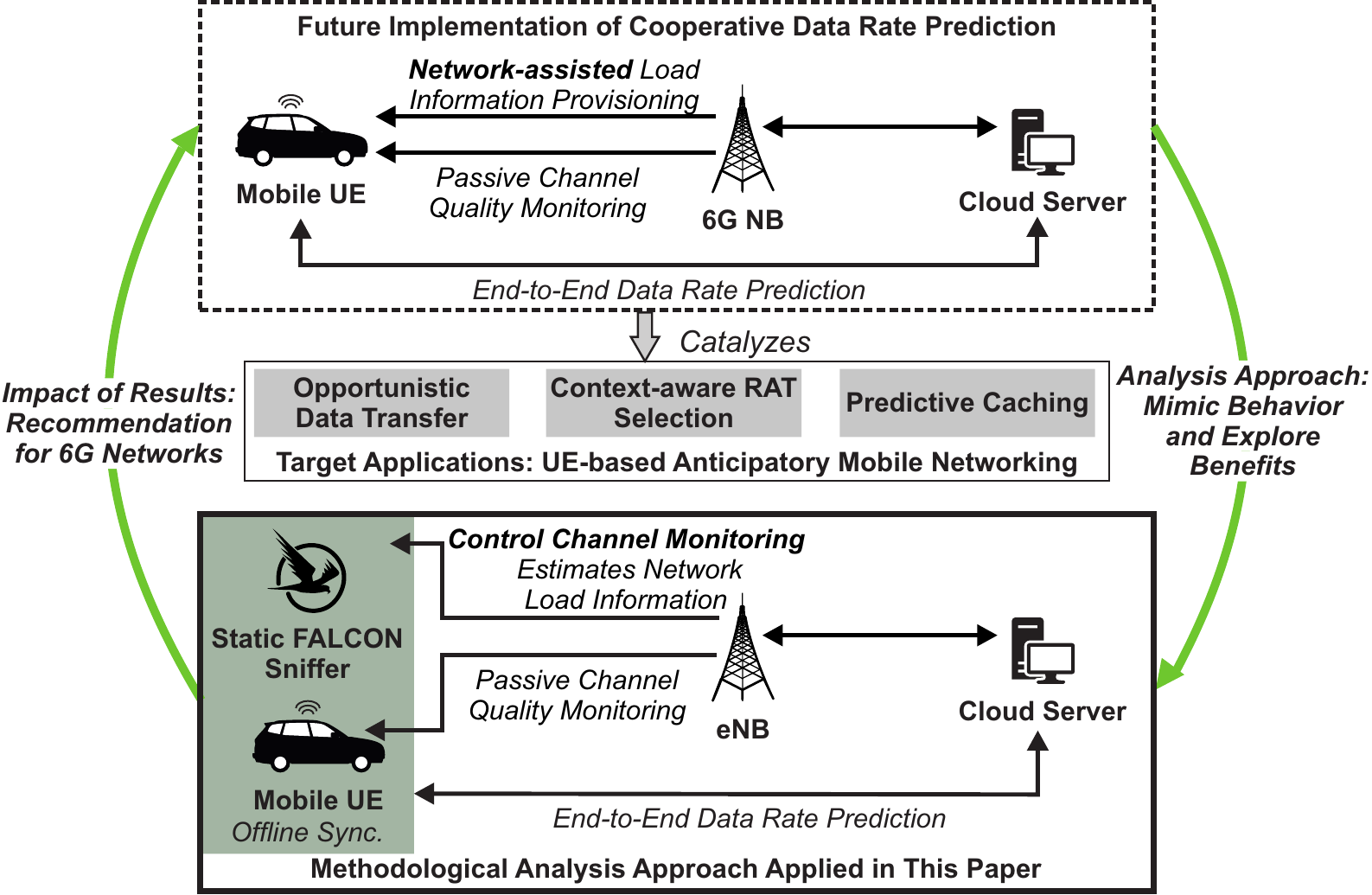}
	\vspace{-0.7cm}
	\caption{Overview of the proposed cooperative data rate prediction approach.}
	\label{fig:scenario}
\end{figure}

%
% Drawbacks of pure client-based methods
%
For moving vehicles, the prediction of the currently achievable end-to-end data rate is a challenging task. Different studies (see Sec.~\ref{sec:related_work}) have shown that network quality information can serve as a meaningful indicator for throughput prediction. However, the main drawback of pure \client prediction approaches is their unawareness of the potentially available network resources and the traffic load related to other active users.
In this paper, we explore the benefits of \emph{cooperative} data rate prediction as a possible method deployed in future 6G networks where network load information could be announced via control or broadcast channels. An overview of the proposed approach and the research goals is illustrated in Fig.~\ref{fig:scenario}.
%
%
%
%We combine \client measurements with an \ac{SDR}-based \ac{FALCON} \cite{Falkenberg/Wietfeld/2019a} sniffer for passive control channel analysis in order to mimic the expected behavior of such a system.
We mimic such a system by combining mobile \ac{UE} measurements with information about the cell-wide radio resource allocations which are revealed by analysis of \ac{PDCCH} using the \ac{SDR}-based \ac{FALCON}~\cite{Falkenberg/Wietfeld/2019a} sniffer.

%
% Structure of the paper
%
The remainder of the paper is structured as follows. After discussing the related work in Sec.~\ref{sec:related_work}, we present the proposed cooperative prediction approach in Sec.~\ref{sec:approach}. Afterward, the applied methodology for the proof-of-concept evaluation is introduced in Sec.~\ref{sec:methods} and finally, the results of the real world performance analysis are presented and discussed in Sec.~\ref{sec:results}.

%
% Fig. Scenario
%
%
%
% !TeX spellcheck = en_US
\section{Related Work} \label{sec:related_work}

%
% Tab. Prediction
%
\begin{table}[]
	%\footnotesize
	\centering
	\caption{Context Domains of Existing Data Rate Prediction Approaches}
	\begin{tabular}{lcccc}
		\toprule
		\textbf{Study} & \textbf{Channel} & \textbf{Load} & \textbf{Mobility} & \textbf{Application} \\
		\midrule
		%
		% Samba/etal/2017a: RSRP, RSRQ, f, indoor/outdoor, Distance to cell, Speed, Average Number of Users
		%
		Samba \cite{Samba/etal/2017a} & \full & \missing & \full & \missing \\
		%
		% Jomrich/etal/2018a: Signal Strength Level, Timing Advance, RSRP, RSRQ, CQI, RSSNR, average speed of vehicle
		%
		Jomrich \cite{Jomrich/etal/2018a} & \full & \partly & \full & \missing \\
		%
		% Wei/etal/2018a: RSSI, Cell Id, Lat, Lon, Orientation, Acceleration -> RNN
		%
		Wei \cite{Wei/etal/2018a} & \full & \missing & \full & \missing \\
		%
		% Cainey/etal/2014a: RSRP, RSRQ, SINR -> Linear Regression
		%
		Cainey \cite{Cainey/etal/2014a} & \full & \partly & \missing & \missing \\
		%
		% Akselrod/etal/2017a: SINR, RSSI ->
		%
		Akselrod \cite{Akselrod/etal/2017a} & \full & \missing & \missing & \missing \\
		%
		% Riihijarvi/Mahonen/2018a: C/I, RSS, Velocity -> Random Forest, ANN, Linear Regression
		%
		Riihijarvi \cite{Riihijarvi/Mahonen/2018a} & \full & \missing & \full & \missing \\
		%
		% Herrera-Garcia/etal/2019a:
		%
		%Herrera-Garcia \cite{Herrera-Garcia/etal/2019a}$^1$ & \full & \full & \missing & \full \\
		%
		% Sliwa/Wietfeld/2019b:
		%
		Sliwa \cite{Sliwa/Wietfeld/2019b} & \full & \partly & \full & \full \\
		Falkenberg \cite{Falkenberg/Wietfeld/2017c} & \full & \full & \missing & \partly \\
		\midrule
		\textbf{This paper}  & \full & \full & \full & \full \\

		\bottomrule
		
	\end{tabular}
	
	\vspace{0.1cm}
	\full~Full consideration, \partly~Partial consideration, \missing~No consideration
		
	\label{tab:data_rate_prediction}
\end{table}

%
%
%

%\cite{Ali/etal/2019a}

%
% DATA RATE PREDICTION
%
\textbf{Data rate prediction} in vehicular networks is a highly challenging task due to the complex interdependency between mobility-, channel-, and network-dependent factors. As the resulting dimensionality of the problem is typically too complex for analytical approaches, machine learning models that implicitly consider hidden interdependencies between measurable variables are applied. A methodological summary of machine learning for wireless communications is provided by \cite{Ye/etal/2018a}. 
%
% Supervised Learning
%
Data rate prediction can be considered as a \emph{regression} task where \emph{supervised} learning is applied to train a predictor $f$ on measurement data $\mathbf{X}$ \emph{labeled} with ground truth values $\mathbf{Y}$ such that $f:\mathbf{X}\rightarrow\mathbf{Y}$. After the training phase, the model can be utilized to make predictions $\mathbf{\widetilde{Y}}$ on unlabeled data.
%
% Active vs Passive
%
A distinction is made into \emph{active} and \emph{passive} prediction methods. While the former apply time series analysis on continuously monitored data rate measurements of ongoing transmissions, the latter only consider passively measurable signal quality indicators (e.g., the \ac{SINR}) without any ongoing transmission.
Since active approaches introduce additional traffic to the network, this paper focuses on the passive prediction approach which is also studied by the authors of \cite{Sliwa/Wietfeld/2019b, Samba/etal/2017a, Jomrich/etal/2018a, Wei/etal/2018a, Cainey/etal/2014a, Akselrod/etal/2017a, Riihijarvi/Mahonen/2018a}. The main conclusions of the previous studies are summarized as follows:
\begin{itemize}
	%
	% Passive indicators
	%
	\item All considered evaluations agree that passively measurable network quality indicators \cite{3GPP/2016a} are highly correlated to the data rate and can be used to forecast the achievable throughput.
	%
	% Payload size
	%
	\item As discussed in \cite{Sliwa/Wietfeld/2019b}, the prediction accuracy highly depends on the payload size of the to be transmitted data packet. Integrating the latter into the prediction process allows to implicitly consider effects such as the \ac{TCP} slow start as well as cross-layer dependencies between the transport layer and the channel coherence time.
	%
	% Prediction Models
	%
	\item In most evaluations (e.g., \cite{Sliwa/Wietfeld/2019b, Samba/etal/2017a, Jomrich/etal/2018a}), the highest prediction accuracy is achieved by \ac{CART}-based models such as \acp{RF}. More complex methods like \emph{deep learning} suffer from the limited amount of available training data.
\end{itemize}
%
%
%

%
% Fig. Approach
%
\begin{figure}[tb]  	
	%\vspace{-0.5cm}
	\centering		  
	\includegraphics[width=1\columnwidth]{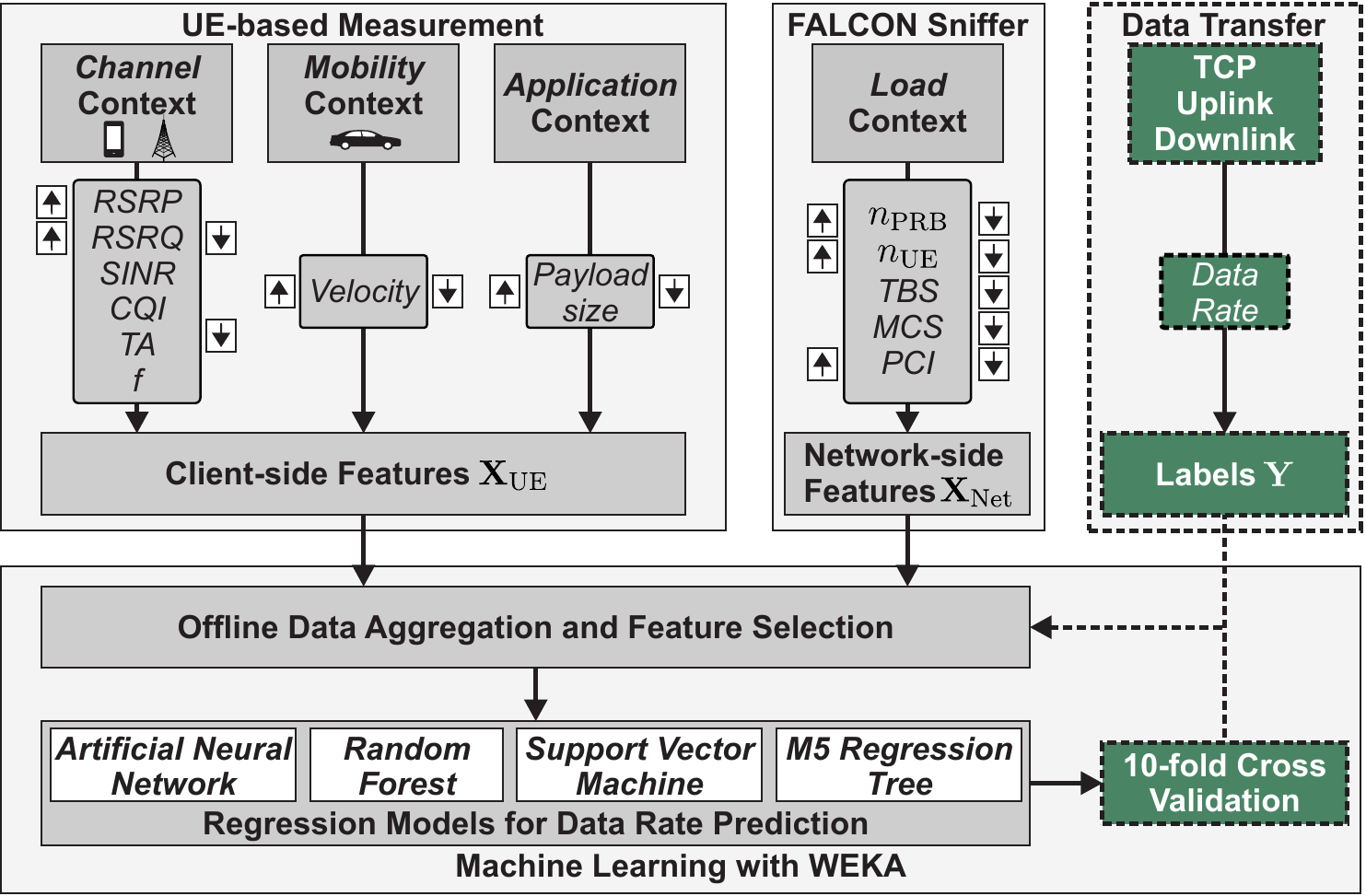}
	\vspace{-0.5cm}
	\caption{System architecture model for the proposed cooperative data rate prediction approach. Relevant features after feature selection for up- and downlink are indicated by arrows \boxedArrowUp and \boxedArrowDown, respectively.}
	\label{fig:cooperative_prediction}
	\vspace{-0.7cm}
\end{figure}
%
%
%

%
% CONTROL CHANNEL ANALYSIS
%
\textbf{Control channel analysis} is an accurate method for fine-grained monitoring of the overall cell activity, which is invisible to conventional \acp{UE}.
In contrast to the mere observation of the spectral power density, which only provides a statement about the total utilization of available resources, the control channel analysis additionally breaks down the number of all competing subscribers as well as their individual throughputs.
In terms of spectral radio resources, this enables a prediction of the \emph{pie piece size} a new cell user shall expect for an intended transmission.
In \ac{LTE} networks, individual resource allocations for uplink and downlink, namely \ac{DCI}, are signaled via \ac{PDCCH} to the \acp{UE} once per millisecond.
Although this information is not encrypted, it is not readily accessible to an observer, since the integrity check presupposes knowledge of all \acp{RNTI} of currently active \acp{UE}.
However, \acp{RNTI} are assigned only once during initial random access response and may take place prior to the observation period or even on a different component carrier in case of carrier aggregation.
Therefore, real time cell monitoring requires advanced methods for discovery of missed \ac{RNTI} assignments and reliable \ac{DCI} validation techniques.
To the best of our knowledge, \ac{FALCON}~\cite{Falkenberg/Wietfeld/2019a} is currently the most accurate open source instrument for performing this task which reliably discloses currently active \acp{RNTI} and reveals all \ac{DCI} from \ac{PDCCH}.
In order to not miss \acp{DCI} that are addressed to cell-center users and include less redundancy for error correction, the \ac{FALCON} sniffer must be placed in proximity of the antenna of the monitored cell.
Consequently, data rate predictions based on cell load, such as \cite{Falkenberg/Wietfeld/2017c}, are bound to stationary scenarios.
%
% Static environments: No handovers, place the sniffer at a location with good network quality
%
%

% Novel Approach
%
The motivation of this paper is to bring together stationary cell load information with measurements of mobile \acp{UE}, hence compensating the drawbacks of both approaches. As a result, a unique level of considered context domains is achieved and can be exploited for mobile data rate prediction. Tab.~\ref{tab:data_rate_prediction} summarizes the considered context domains exploited by related studies on the proposed cooperative approach.

% !TeX spellcheck = en_US

%
% Fig. Approach
%
%\begin{figure}[tb]  	
%	%\vspace{-0.5cm}
%	\centering		  
%	\includegraphics[width=1\columnwidth]{fig/cooperative_prediction}
%	%\vspace{-0.5cm}
%	\caption{System architecture model for the proposed cooperative data rate prediction approach. Relevant features after feature selection for up- and downlink are indicated by arrows \boxedArrowUp and \boxedArrowDown, respectively.}
%	\label{fig:cooperative_prediction}
%	%\vspace{-0.5cm}
%\end{figure}
%%
%%
%%

\section{Cooperative Data Rate Prediction}  \label{sec:approach}

In this section, the proposed cooperative data rate prediction approach is presented.
Based on the overall system architecture model shown in Fig.~\ref{fig:cooperative_prediction}, the different components are explained.

%
% DATA ACQUISITION
%
\textbf{Hybrid data acquisition}: Different features for the machine learning process are captured by the mobile \ac{UE} and the static \ac{FALCON} sniffer.
%
% UE
%
The mobile \ac{UE} determines features from multiple context domains which are brought together in the client-side feature set $\mathbf{X}_{\text{UE}}$.
These comprise
%
% Client-based Features
%
\begin{itemize}
	\item \textbf{Channel context}: \ac{RSRP}, \ac{RSRQ}, \ac{SINR}, \ac{CQI}, \ac{TA} and the carrier frequency $f$;
	\item \textbf{Mobility context}: Velocity of the vehicle;
	\item \textbf{Application context}: Payload size of the intended transmission.
\end{itemize}
In the training phase, the resulting throughput of each data transmissions is used as the label $\mathbf{Y}$ for the prediction process.

%
% FALCON
%
In parallel, a statically deployed \ac{FALCON} sniffer captures the cell's \textbf{load context} to derive network-side features $\mathbf{X}_{\text{Net}}$ from the monitored resource allocations.
These comprise, separated by up- and downlink, the statistics (i.e. average and standard deviation) of the number of active users $n_{\text{UE}}$, the number of assigned \acp{PRB} $n_{\text{PRB}}$, \ac{MCS}, and \ac{TBS}, within a single subframe and within an observation window of \SI{1}{\second}.
%
% Network-based Features
%
%
%
% Offline Sync
%
Based on synchronized timestamps, the data sets $\mathbf{X}_{\text{UE}}$ and $\mathbf{X}_{\text{Net}}$ aggregated offline for the training.

%
% FEATURE SELECTION
%
\textbf{Feature selection}: In order to maximize the information gain, redundancies related to highly correlated input variables are removed by an iterative feature selection for each transmission direction.
The algorithm is initialized with an empty feature set $\mathbf{X} = \emptyset$ and a feature candidate set $\mathbf{C} = (\mathbf{X}_{\text{UE}} \cup \mathbf{X}_{\text{Net}})\setminus \mathbf{X}$ of all remaining features. 
In each iteration, the performance $R^2_i$ (cf. Sec.~\ref{sec:methods_ml}, Eq.~\ref{eq:r2}) is evaluated by including one additional feature $c_{i} \in \mathbf{C}$ for all $i \in [1, |\mathbf{C}|]$ and $\mathbf{X}$ is appended by the best feature $c_{m}$ with $m = \argmax_{i\in [1, |\mathbf{C}|]}(R^2_i)$.
%
%In each iteration, the performance gain -- evaluated based on the criteria defined in Sec.~\ref{sec:methods_ml} -- achieved by adding one additional feature $c_{i} \in \mathbf{C}$ is analyzed for all $i \in [1, |C|]$ and the best feature candidate $c_{\max}$ is added to $\mathbf{X}$.  
%
%
%
The algorithm terminates as the prediction performance $R^2_i$ decreases or $\mathbf{C} = \emptyset$.
For a feature set of length $n = |\mathbf{C}|$, up to $\frac{n(n-1)}{2}$ models are trained and evaluated.
%
%
% Resulting Features:
% ul: payload;avg_nof_prb_per_UE_ul;avg_nof_prb_per_subframe_ul;rsrp;std_nof_prb_per_UE_ul;tot_nof_UE_ul;speed;pci;tot_nof_prb_ul;rsrq
%
%
% dl: payload;avg_mcs_dl;tot_tbs_dl;tot_nof_prb_dl;std_nof_UE_per_subframe_dl;speed;std_nof_prb_per_UE_dl;pci;rsrq;ta;std_nof_prb_per_subframe_dl;avg_tbs_dl;f;rsrp;avg_nof_prb_per_subframe_dl;std_mcs_dl;avg_nof_prb_per_UE_dl
In Fig.~\ref{fig:cooperative_prediction}, the features selected by the algorithm for up- and downlink data rate prediction are indicated by arrows \boxedArrowUp and \boxedArrowDown, respectively.

%
% MACHINE LEARNING
%
\textbf{Machine learning models}: The actual prediction is carried out with multiple supervised machine learning models. Parameters are chosen based on grid search in a preprocessing step.
%
% Models
%
\begin{itemize}
	\item \textbf{\acf{ANN}} \cite{LeCun/etal/2015a} with two hidden layers consisting 15 nodes, momentum $\alpha=0.001$, learning rate $\eta=0.1$ and 500 epochs. 
	\item \ac{CART} methods \textbf{\acf{RF}} \cite{Breiman/2001a} with 100 random trees and \textbf{\acf{M5}} \cite{Quinlan/1992a}.
	\item \textbf{\acf{SVM}} \cite{Cortes/Vapnik/1995a} with \ac{RBF} kernel which is trained with \ac{SMO}.
\end{itemize}
% !TeX spellcheck = en_US

\begin{figure}[]  	
	%\vspace{-0.5cm}
	\centering		  
	\includegraphics[width=1.0\columnwidth]{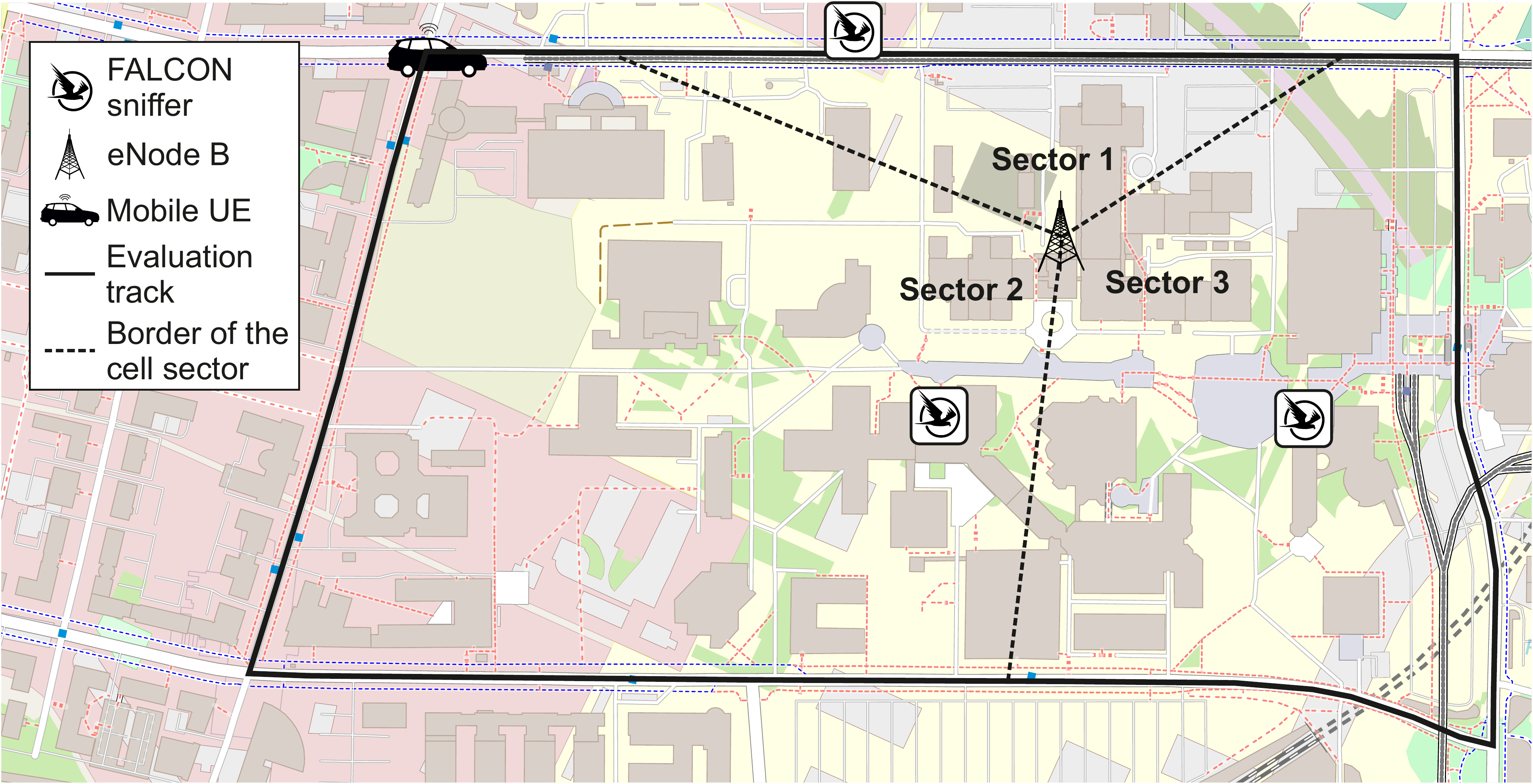}
	%\vspace{-0.5cm}
	\caption{Trajectory of the drive tests in a campus region including a serving cell with three sectors and locations of the associated \ac{FALCON} sniffers (Map data: ©OpenStreetMap contributors, CC BY-SA).}
	\label{fig:map}
	%\vspace{-0.5cm}
\end{figure}
%
%
%

%
% Fig. RMSE Boxplots
%
\begin{figure}[]  	
	%\vspace{-0.5cm}
	\centering		  
	\includegraphics[width=1\columnwidth]{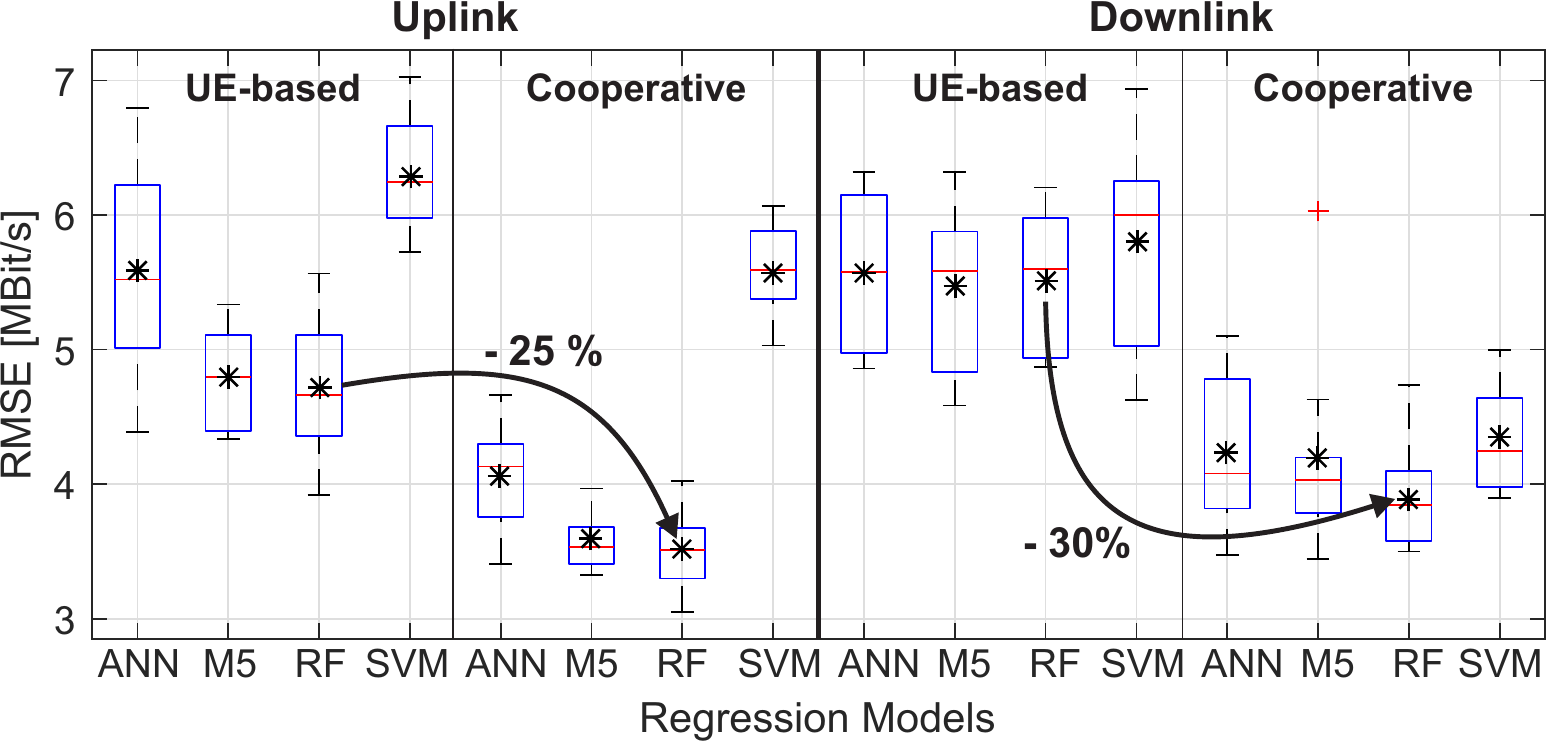}
	\vspace{-0.5cm}
	\caption{Comparison of the resulting predictions errors of \client and cooperative data rate prediction for the considered machine learning models: \acf{ANN}, \acf{M5}, \acf{RF}, and \acf{SVM}. The \client  approach uses the full client-side feature set $\mathbf{X}_{\textbf{UE}}$.}
	\vspace{-0.3cm}
	\label{fig:box_rmse}
\end{figure}

\section{Methodology}  \label{sec:methods}

This section provides an introduction of the methodological setup for the real world data acquisition and the machine learning-based data analysis.

\subsection{Real World Data Acquisition}

%
% Scenario
%
For the proof-of-concept evaluation, drive tests with data transmissions are carried out in a campus region that is covered by three sectors of a single \ac{eNB} that belongs to a public \ac{LTE} network as shown in Fig.~\ref{fig:map}.
%
%
% Android UE
%
The mobile channel quality measurements and the active data transmissions are performed using an off-the-shelf Android-based \ac{UE} (Samsung Galaxy S5 Neo, Model SM-G903F) which executes the measurement application.
\ac{TCP} transmissions are performed periodically each \SI{10}{\second} in uplink and downlink direction through the public \ac{LTE} network. The exchanged payload is chosen randomly in the range of \SI{0.1}{\mega\byte} to \SI{10}{\mega\byte}.

During the drive tests, the entire cell activity is captured by three synchronized \ac{FALCON}\footnote{\ac{FALCON} is available at https://github.com/falkenber9/falcon} sniffers, each placed in one of the base station's sectors in line of sight to the antenna.
Each sniffer comprises a common Laptop running the \ac{FALCON} software and an attached USRP~B210 \ac{SDR} by Ettus Research with a dipole antenna receiving the signal.

%The sniffer is installed within a mobile communication laboratory as shown in Fig.~\ref{fig:photo_crafter} which is parked near the analyzed \acp{eNB}. For the measurements, three static \ac{FALCON} deployments are used in parallel at different locations.

%
% Data set description and Total
%
The considered data set is the result of 92 real world drive tests and consists of measurements for 3027 data transmissions.
It includes measurements during peak noon, while the mobile network is very congested, as well as measurements during the night, with almost no activity by other participants.

\subsection{Machine Learning-based Data Analysis} \label{sec:methods_ml}

%
% TOOL
%
All data analysis tasks are carried out with \ac{WEKA} \cite{Hall/etal/2009a}. In order to avoid overfitting, we apply 10-fold cross validation and analyze the statistical derivations between the different folds.
%
% QUALITY MEASURES
%
For assessing the quality of the data rate prediction, we consider multiple typical quality measures for regression tasks.
The \emph{coefficient of determination} $R^2$ is a statistical metric widely used by the related work. It describes the amount of response variable derivation that is explained by the trained regression model and is calculated as
%
% Eq. R^2
%
\begin{equation}
R^{2} = 1- \frac{\sum_{i=1}^{N}\left(\tilde{y}_{i} - y_{i} \right)^{2}}{\sum_{i=1}^{N}\left(\bar{y} - y_{i} \right)^{2}}\label{eq:r2}
\end{equation}
with the current prediction $\tilde{y}_{i}$, the current label $y_{i}$, the mean data rate $\bar{y}$, and the number of measurement samples $N$.
In addition, we consider \ac{MAE} and \ac{RMSE}  which are defined as
%
% Eq. MAE
%
\begin{equation*}
	\text{MAE} = \frac{\sum_{i=1}^{N} | \tilde{y}_{i} - y_{i}|}{N},
	\quad
	\text{RMSE} = \sqrt{\frac{\sum_{i=1}^{N} \left(  \tilde{y}_{i} - y_{i} \right)^2}{N}}.
\end{equation*}
%
% Eq. RMSE
%
%\begin{equation}
%	\text{RMSE} = \sqrt{\frac{\sum_{i=1}^{N} \left(  \tilde{y}_{i} - y_{i} \right)^2}{N}}
%\end{equation}
% !TeX spellcheck = en_US

%
% Fig. Scatterplots
%
\newcommand{\sfw}{0.32}
\begin{figure*}[]
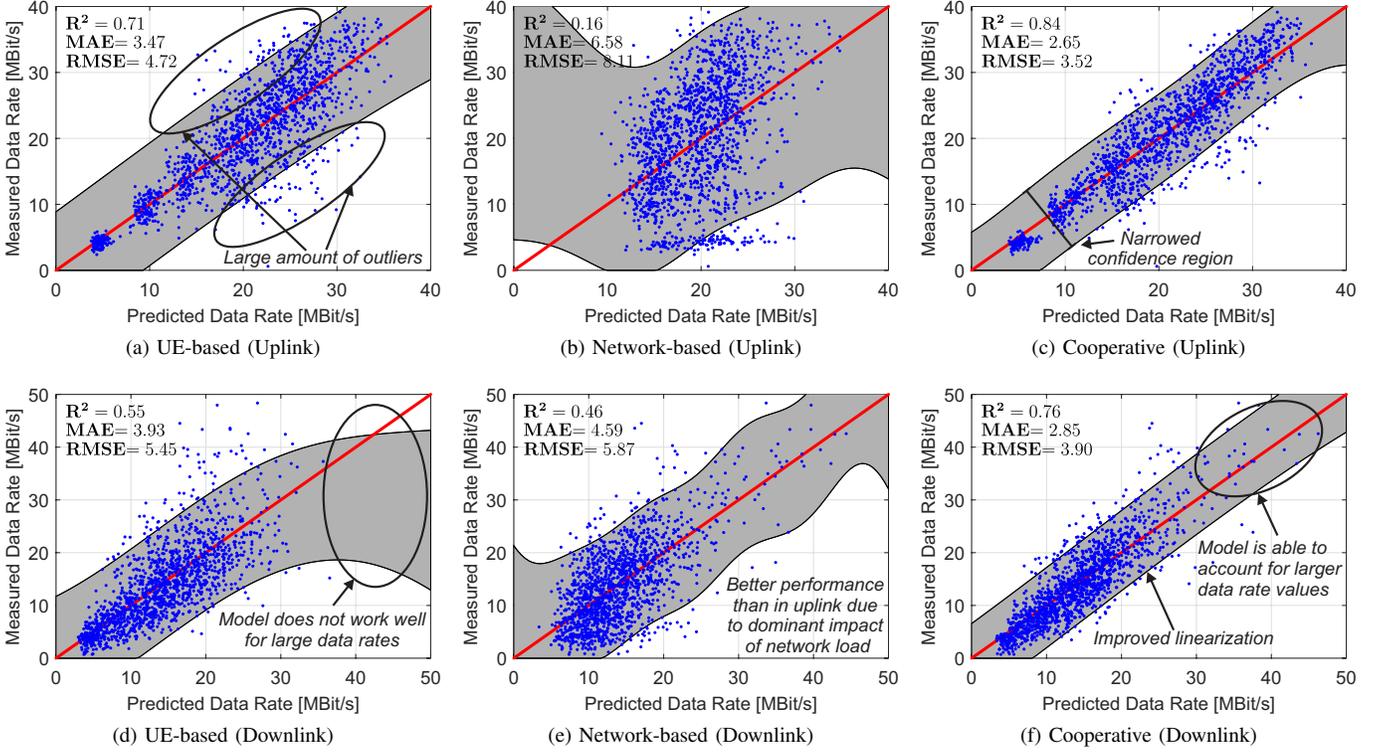
 
	\centering
	
	%
	% UL
	%
	\subfig{fig/client_ul}{\sfw}{\client (Uplink)}
	\subfig{fig/net_ul}{\sfw}{\net (Uplink)}
	\subfig{fig/coop_ul}{\sfw}{\coop (Uplink)}

	%
	% DL
	%
	\subfig{fig/client_dl}{\sfw}{\client (Downlink)}
	\subfig{fig/net_dl}{\sfw}{\net (Downlink)}
	\subfig{fig/coop_dl}{\sfw}{\coop (Downlink)}
	
	\caption{Performance comparison of different \ac{RF} data rate prediction approaches in uplink and downlink direction.
		Network-based predictions (b) and (e) only consider cell load information features $\mathbf{X}_{\text{Net}}$ from the \ac{FALCON} measurements.
		The gray area shows the 0.95 confidence area derived by applying GPR on the results of the prediction model. Diagonal lines illustrate perfect predictions. \ac{MAE} and \ac{RMSE} are expressed in MBit/s.}
	\label{fig:model_performance}
	%\vspace{-0.5cm}
\end{figure*}

\section{Results of the Real World Proof-of-Concept Evaluation}  \label{sec:results}

In this section, we present and discuss the results of the real world proof-of-concept evaluation. 
%
% Feature Sets for Cooperative Approach
%
At first, the achievable prediction performance is compared for different machine learning models for the \client and cooperative approaches. Fig.~\ref{fig:box_rmse} shows the resulting \ac{RMSE} values.
%
% UE-based vs Cooperative
%
It can be seen that the cooperative approach is able to reduce the average \ac{RMSE} by up to 25\% in uplink and 30\% in downlink transmission direction by considering features that indicate the current network load.
%
% Random Forest
%
In all variants, the lowest \ac{RMSE} is achieved by the \ac{RF} model which is in consensus with related work \cite{Sliwa/Wietfeld/2019b, Samba/etal/2017a, Jomrich/etal/2018a}.
%
% M5
%
However, it is remarkable that \ac{M5} achieves an almost comparable performance level, as the model is far less complex than the \ac{RF}.
%
% Models behave more similar in the Downlink direction -> Dominante Network Features?
%
In the downlink direction, the machine learning models show a more consistent behavior. As discussed in the following paragraph, the traffic load has a dominant impact on the resulting data rate which results in a more linear relationship between the considered features.

Based on these observations, the behavior of the \ac{RF} model is further investigated. Fig.~\ref{fig:model_performance} shows the resulting prediction performance for \client, network-based, and the proposed cooperative approach. The statistical behavior of the model is further illustrated with a confidence region derived by applying \ac{GPR} on the prediction results.
%
% Pure UE-based
%
It can be seen that the \client approach achieves a generally good correlation between predictions and measurements. It should be denoted that this methodological approach only requires an off-the-shelf \ac{UE} to perform the measurements, which means a lower hardware-related effort than the \ac{SDR}-enabled cooperative approach. However, large outliers occur due to missing network load information.
%
% Pure Network-based
%
Pure network-based prediction is unaware of the radio channel conditions of the targeted mobile \ac{UE} and only able to consider the current network load.
%
% Uplink vs Downlink
%
As the downlink is typically more congested than the uplink \cite{Bui/etal/2017a}, the achievable data rate is more determined by the network-related than the channel-related features. Therefore, the prediction works more reliable in the downlink direction.
%
% Cooperative
%
In both transmission directions, the proposed cooperative approach is able to compensate the major limitations of the individual approaches. As a result, the error spread is significantly reduced which results in a more linear and more narrow confidence interval.

% !TeX spellcheck = en_US
\section{Conclusion}

%
% Introduction
%
In this paper, we presented a cooperative approach for cellular data rate prediction which brings together \client channel quality sensing with network-based load estimation in vehicular scenarios.
%
% Solution appraoch
%
In order to mimic the possible behavior of network-assisted throughput prediction in future 6G networks, a real world performance evaluation based on the \ac{FALCON} sniffer was conducted in a vehicular context.
%
% Results
%
It was shown that \ac{SDR}-based approaches are capable of extracting network load information based on control channels analysis and that this knowledge can be utilized to significantly improve the data rate prediction accuracy for mobile \acp{UE} in both transmission directions. 
%
% Future work
%
Within 5G networks, data analytics-based (e.g., \ac{NWDAF}-enabled) network optimization is currently solely considered for the network infrastructure side. Although it is not clear which kind of intelligence future 6G networks will implement, we strongly advertise that the obtained traffic load information should be actively shared with the \acp{UE} in order to catalyze network-assisted \client optimization methods.
%
% Synchronization -> Reduced Hardware effort
%
This way, further enhancements of the prediction accuracy can be expected as the need for synchronizing multiple time series measurements is removed.

% !TeX spellcheck = en_US
\ifdoubleblind

\else
\section*{Acknowledgment}

\footnotesize
Part of the work on this paper has been supported by Deutsche Forschungsgemeinschaft (DFG) within the Collaborative Research Center SFB 876 `Providing Information by Resource-Constrained Analysis', projects A4\,+\,B4.
\fi

\bibliographystyle{IEEEtran}
\bibliography{Bibliography}

\end{document}